\documentclass[aps,prl,twocolumn,groupedaddress,showpacs]{revtex4}

\usepackage{graphicx}
\usepackage{amsmath}
\usepackage{amssymb}

\begin{document}
\title{The critical temperature of a trapped, weakly interacting Bose gas}
\author{F.\ Gerbier}
\email[email: ]{fabrice.gerbier@iota.u-psud.fr}
\author{J.\ H.\ Thywissen}
\altaffiliation[current address: ]{Department of Physics, University of
Toronto,  Toronto, ON, M5S 1A7, Canada.}
\author{S.\ Richard}
\author{M.\ Hugbart}
\author{P.\ Bouyer}
\author{A.\ Aspect}
\affiliation{Groupe d'Optique Atomique, Laboratoire Charles Fabry
de l'Institut d'Optique \footnote{UMR 8501 du CNRS}, 91403 Orsay
Cedex, France}
%
\date{\today}
\begin{abstract}
We report on measurements of the critical temperature of a
harmonically trapped, weakly interacting Bose gas as a function of
atom number. Our results exclude ideal-gas behavior by more than
two standard deviations, and agree quantitatively with mean-field
theory. At our level of sensitivity, we find no additional shift
due to critical fluctuations. In the course of this measurement,
the onset of hydrodynamic expansion in the thermal component has
been observed. Our thermometry method takes this feature into
account.
\end{abstract}
 \maketitle

Degenerate atomic Bose gases provide an ideal testing ground for
the theory of quantum fluids. First, their diluteness makes
possible  first-principles theoretical approaches
\cite{dalfovo1999a}. Second, thanks to the powerful experimental
techniques of atomic physics, static and dynamic properties can be
studied quantitatively through a wide range of temperature and
densities. Furthermore, the inhomogeneity induced by the external
trapping potential leads to entirely new behavior, when compared
to bulk quantum fluids.

Atomic interactions have previously been found to affect deeply
the dynamical behavior of trapped Bose gases at finite
temperatures \cite{jin1997a,marago2001a}. By contrast, the
influence of interactions on {\em thermodynamics} is less
pronounced \cite{giorgini1996a}, and has been less studied
experimentally. Pioneering work on thermodynamics
\cite{ensher1996a} concentrated essentially on the ground state
occupation, and the role of interactions was somewhat hidden by
finite size effects \cite{dalfovo1999a}. Though several such
measurements have been reported \cite{marago2001a,mewes1996a}, to
our knowledge a decisive test of the role of interactions is still
lacking. In this Letter, we focus on the critical temperature
$T_{\rm c}$ of a harmonically trapped $^{87}$Rb Bose gas to
demonstrate the influence of interactions on the thermodynamics.
We study the behavior of $T_{\rm c}$ as a function of the number
of atoms at the transition, for a fixed trapping geometry. We find
a deviation from ideal-gas behavior, towards lower critical
temperatures, whose signification will be discussed below. In the
course of this study, we have observed that collisions induce an
anisotropy in the free expansion of the cloud even far from the
hydrodynamic regime \cite{kagan1997a,schvarchuck2002a}. We correct
for this effect in our temperature measurement.

For an ideal Bose gas in a harmonic trap, the critical temperature
is \cite{dalfovo1999a}
\begin{equation}\label{tcideal} k_{\rm B} T_{\rm c}^{\rm ideal} =
\hbar \bar{\omega}\Big{(}\frac{N}{\zeta{(3)}} \Big{)}^{1/3} -
\frac{\zeta(2)}{6\zeta(3)}\hbar (\omega_{z }+2\omega_{\perp}).
\end{equation}
The first term on the right hand side is the transition
temperature $T_{\rm c}^0$ in the thermodynamic limit, and the
second represents finite-size corrections. Here $N$ is the total
atom number, $\omega_{\perp}$ and $\omega_{z}$ are the trapping
frequencies, $\bar{\omega}=\omega_{\perp}^{2/3}\omega_{ z}^{1/3}$
is their geometrical average, and $\zeta$ is the Riemann zeta
function.

As stated earlier, the main goal of this paper is to probe the
role of two-body repulsive interactions on $T_{\rm c}$.
Corrections to the ideal gas formula depend on the ratio between
the s-wave scattering length $a$ and $\lambda_{\rm 0}=h/\sqrt{M
k_{\rm B} T_{\rm c}^0}$, the de Broglie wavelength at the
transition ($\lambda_0=(2\pi)^{1/2}~\sigma [N/\zeta(3)]^{-1/6}$
for a harmonically trapped gas, where
$\sigma=\sqrt{\hbar/M\bar{\omega}}$ is the mean ground state
width). In a trapped gas, the dominant effect of interactions can
be understood using a simple mean-field picture
\cite{giorgini1996a}: interactions lower the density in the center
of the trap $n({\bf 0})$, and accordingly decrease the temperature
$T$ that meets Einstein's criterion $n({\bf 0})\lambda_{\rm
0}^3=\zeta(3/2)$. The magnitude of this reduction has been
calculated to leading order \cite{giorgini1996a},
\begin{equation}\label{shift}
\frac{T_{\rm c}-T_{\rm c}^{\rm ideal}} {T_{\rm c}^{0}} = -a_1
\frac{a}{\lambda_0} \approx - 1.326 \frac{a}{\sigma} N^{1/6},
\end{equation}
where $a_1\approx3.426$ \cite{giorgini1996a,arnold2001a}. In this
work, the finite-size correction in Eq.~(\ref{tcideal}) changes
$T_{\rm c}^0$ by at most 2\%, whereas the interactive shift
(\ref{shift}) can be as high as 10\%. The measurements presented
below are in quantitative agreement with the prediction of Eq.
(\ref{shift}).

In addition, as dicussed in
\cite{arnold2001a,houbiers1997a,holzmann2003a,gruter1997a,baym,Tc_mc},
critical fluctuations that develop in the system near $T_{\rm c}$
are expected to favor the formation of the condensate and thus to
increase $T_{\rm c}$. In the case of a uniform Bose gas
\cite{gruter1997a,baym,Tc_mc}, this is the leading effect, because
the critical temperature is not affected at the mean-field level.
The correction $\delta T_{\rm c}/T_{\rm c}^0 = + c_1 a/\lambda_0$,
with $c_1\approx1.3$ \cite{Tc_mc}, can be traced back to density
fluctuations with wavelength much larger than the correlation
radius $r_{\rm c} \sim \lambda_0^2/a$ \cite{baym}. This upwards
trend, which has been observed experimentally in a dilute sample
of $^4$He adsorbed in a porous glass \cite{reppy2000a}, is quite
sensitive to the presence of an external potential
\cite{arnold2001a,holzmann2003a}. In the harmonically trapped case
of interest here, the contribution of long wavelength excitations
to the shift in $T_{\rm c}$ scales as a higher power of
$a/\lambda_0$, making it negligible when compared to the
``compressional'' shift given by Eq. (\ref{shift}). The
quantitative agreement we find with the mean-field result can be
considered evidence of this effect, and highlights the important
role played by the trapping potential.

Our experimental setup to reach Bose-Einstein condensation in the
$| F=1; m_{\rm F}=-1 \rangle$ hyperfine ground state of $^{87}$Rb
is similar to that used in \cite{richard2003a}. The trapping
frequencies are $\omega_{\perp}/2\pi=413(5)$\,Hz and
$\omega_{z}/2\pi=8.69(2)$\,Hz in the present work. To reduce
non-equilibrium shape oscillations that occur in such anisotropic
traps upon condensation \cite{richard2003a,schvarchuck2002a}, the
last part of the evaporation ramp is considerably slowed down (to
a ramp speed of $200$\,kHz/s) and followed by a $1$\,s hold time
in the presence of a radio-frequency shield. We ensure good
reproducibility of the evaporation ramp in the following way. We
monitor regularly (typically every four cycles) the
radio-frequency $\nu_{0}$ that empties the trap. This allows us to
detect slow drifts of the trap bottom, and to adjust in real time
the final evaporation radio-frequency $\nu_{\rm rf}$ to follow
them. In this way, the ``trap depth'' $\nu_{\rm rf}-\nu_{0}$, is
kept constant within $\pm 2$\,kHz. Since we measure
$\eta=h(\nu_{\rm rf}-\nu_{0})/k_{\rm B} T\approx$ 11 in this final
evaporation stage, we estimate the temperature stability to be
$\pm 10$\,nK.

We infer the properties of the clouds by absorption imaging. After
rapid switch-off of the trap ($1/e$ cut-off time of about
$50$\,$\mu$s), a $22.3$\,ms free expansion, and a repumping pulse,
we probe the ultra-cold cloud on resonance with the $| F=2 \rangle
\rightarrow | F'=3 \rangle$ transition \cite{imaging}. The images
are analyzed using a standard procedure, described for instance in
\cite{Ketterle1999a}. For an ideal thermal cloud above the
transition point, the evolution of the density in time of flight
is related to the initial density profile by simple scaling
relations, so that the column density (integrated along the probe
line-of-sight, almost perpendicular to the long axis of the trap)
is
\begin{eqnarray} \label{g2}
\tilde{n}_{\rm th}({\bf \rho}) & =
& \tilde{n}_{\rm th}({\bf 0}) g_{2}\left\{ \exp
\Big{(}\frac{\mu}{k_{\rm B} T } - \frac{x^{2}}{2 R_{\rm th}^2} -
\frac{z^{2}}{2 L_{\rm th}^2} \Big{)} \right\},
\end{eqnarray}
where $g_{2}(u)=\sum_{j \geq 1}u^{j}/j^{2}$, and $x$ and $z$ are
the coordinates along the tight and shallow trapping axes,
respectively. For mixed clouds containing a normal and a (small)
condensed component, we assume, as usual for a condensate in the
Thomas-Fermi regime \cite{dalfovo1999a}, that one can describe the
bimodal distribution by an inverted parabola on top of an ideal,
quantum-saturated thermal distribution (Eq. (\ref{g2}) with
$\mu=0$). The condensed number $N_0$ is then deduced from
integration of the Thomas-Fermi fit, and the total atom number $N$
from integration over the entire image. We estimate that condensed
fractions as low as 1\% can be reliably detected by the fitting
routine. Absolute accuracy on the value of $N$ and $N_0$ relies on
the precise knowledge of the absorption cross-section of the probe
laser, which depends on its polarization and the local magnetic
field. This cross-section is calibrated by fitting the radial
sizes of condensates with no discernible thermal fraction to the
Thomas-Fermi law $R_0 \propto N_0^{1/5}$, as explained in
\cite{dalfovo1999a,kagan1997a,Ketterle1999a,castin1996a}. We find
a reduction of $4.00(14)$ compared to the reference value
$\sigma_0=3\lambda_{\rm L}^{2}/2\pi$ \cite{imaging}.

\begin{figure}[ht!]
\includegraphics[width=8cm]{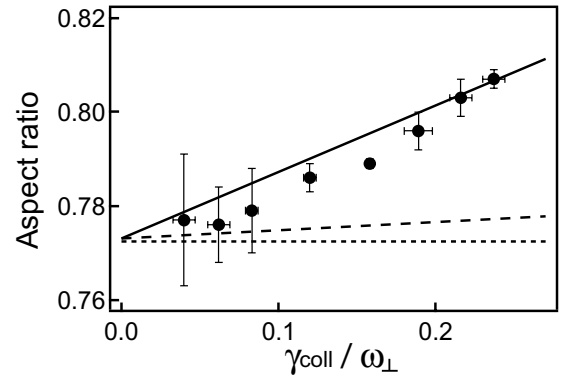}
\caption{Onset of hydrodynamic expansion for trapped clouds above
threshold. Measured aspect ratios after expansion (filled circles,
with statistical error bars) are plotted versus the collision rate
at equilibrium $\gamma_{\rm coll}$. The experimental results are
compared against several hypotheses: a ballistic expansion (dotted
line); a mean-field dominated expansion (dashed line); and a
collisional expansion for a non-condensed Bose gas (solid line).}
\label{thermal}
\end{figure}

We will now discuss the more complex issue of thermometry in some
detail. The temperature is usually inferred from the sizes of the
thermal cloud after a time of flight $t$, assuming a purely
ballistic expansion with isotropic mean velocity,
$v_0=\sqrt{k_{\rm B} T/M}$, as appropriate for an ideal gas. We
show in Fig.~\ref{thermal} that the observed aspect ratio of
non-condensed clouds, in a wide range of temperatures and atom
numbers (corresponding to $1 \lesssim T/T_{\rm c} \lesssim 1.8$),
is actually larger than the value (0.773) expected for an ideal
gas and $\omega_{z}t\approx 1.23$ that corresponds to our
parameters (dotted line in Fig.~\ref{thermal}), in contradiction with
the assumption of an isotropic velocity distribution.

In a very elongated trap, this could be explained by two distinct
collisional effects. First, the initial mean-field energy of the
non-degenerate cloud converts almost completely into {\it
radial} kinetic energy during time-of-fight \cite{menotti2002a},
as for an elongated condensate \cite{kagan1997a,castin1996a}. The
magnitude of this effect is controlled by the ratio $\chi$ of
the mean-field energy to the temperature. In our case, the
parameter $\chi$ does not exceed 0.02, too low to explain the
observed anisotropy (dashed line in Fig.~\ref{thermal}, calculated
along the lines of \cite{menotti2002a}).

Second, as studied theoretically in
\cite{kagan1997a,wu1998a,pedri2003a} and observed in Bose
\cite{schvarchuck2002a} and Fermi gases \cite{fermihydro},
anisotropic expansion occurs for a cloud in the hydrodynamic
regime, {\it i.e.} when the mean free path at equilibrium is
smaller than the dimensions of the sample. In our very elongated
cloud, the mean free path is typically smaller than the axial
length, but much larger than the radial size. Hydrodynamic axial
motion of the thermal particles results in energy transfer from
the axial to the radial degrees of freedom. For weak deviations
from ballistic expansion, this collisional dynamic creates a
velocity imbalance proportional to $\gamma_{\rm coll}$, the
equilibrium collision rate, in agreement with the trend observed
in Fig. \ref{thermal}.

In \cite{pedri2003a}, a set of scaling equations was derived to
investigate how collisions affect the expansion of
a non-condensed cloud. Numerical solution of
these equations, that also include the weak mean field effect,
agrees well with our data (solid line in Fig.~\ref{thermal}). The calculation makes use of the results of
\cite{kavoulakis2000a} for the collision rate of a non-condensed,
almost ideal Bose gas, in general larger (by as much as 70\% close
to $T_{\rm c}$) than the classical collision rate with the same
$N$ and $T$ \cite{gammacoll}. In view of the satisfactory
agreement of our data with the scaling theory, we conclude that
the observed anisotropy is a signature of the onset of
hydrodynamic expansion.

In the regime $\gamma_{\rm coll }\lesssim \omega_{\perp}$, where
the anisotropy is weak and increases linearly with $\gamma_{\rm
coll}$, kinetic energy conservation suggests that mean square
expansion velocities take the form $\langle v_{x}^2 \rangle /v_0^2
\approx 1+\beta \gamma_{\rm coll}/2\omega_\perp$, and $\langle
v_{z}^2 \rangle/ v_0^2\approx1-\beta\gamma_{\rm
coll}/\omega_\perp$, where $\beta$ depends in general on
$\omega_\perp,\omega_z,t$ (fixed for the measurement presented
here). These simple forms are confirmed by the numerical solution
described above. Provided the expansion velocities along both axes
are measured, they allow to infer the initial mean square velocity
$v_0$ and temperature $T$, independently of the coefficient
$\beta$ \cite{temp}. Were this correction not applied, a
systematic 10-15\% discrepancy between the axial and radial
temperature would remain. Note that mean-field effects are not
corrected by this procedure. As stated above, they change the
expansion energy by 2\% at most. This error is not significant
when compared to calibration uncertainties, that limit the
accuracy of $T$ to about 5\% \cite{errT}.

\begin{figure}[ht!]
\includegraphics[width=8cm]{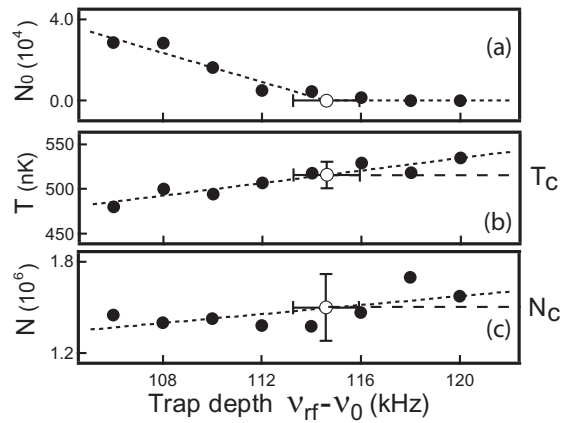}
\caption{Procedure to locate the transition point. We plot the
condensed number (a), temperature (b), and total atom number (c)
as a function of the trap depth, fixed by the final rf frequency
$\nu_{\rm rf}$ and the trap bottom $\nu_0$. The transition point,
shown as a hollow circle  (with statistical error bars), is found
from a fit [dotted curve in (a)], and reported in (b) and (c) to
find $T_{\rm c}$ and $N_{\rm c}$.} \label{findtc}
\end{figure}

Having identified an appropriate thermometric technique, we turn
to the measurement of the critical temperature as a function of
atom number. Data were taken in a narrow range around $T_{\rm c}$.
From the two-component fit, we extract the number of condensed
atoms, the temperature, and the total atom number as a function of
the trap depth, as shown in Fig.~\ref{findtc}a, b and c,
respectively. The trap depth at which the transition point is
reached, $(\nu_{\rm rf}-\nu_0)_{\rm c}$, is taken to be the point
at which a linear fit to the condensed number data crosses zero (a
linear approach towards $T_{\rm c}$ is consistent with the
simulations reported in \cite{krauth1996a}). The temperature and
total number are also fitted assuming a linear dependency on
$\nu_{\rm rf}$, and from the value $(\nu_{\rm rf}-\nu_0)_{\rm c}$,
we extract the critical atom number $N_{\rm c}$ and critical
temperature $T_{\rm c}$.

\begin{figure} [ht!]
\includegraphics[width=8cm]{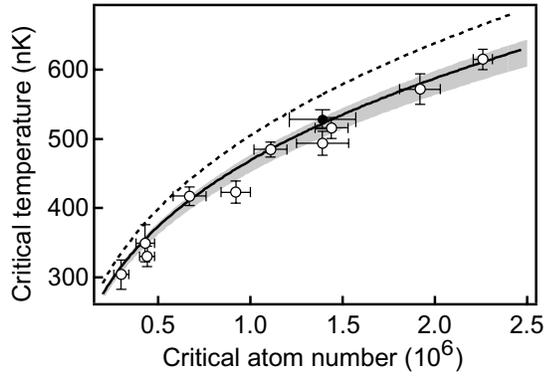}
\caption{Critical temperature as a function of atom number at the
transition. The experimental points (circles) are lower than the
ideal gas law Eq.~(\ref{tcideal}) (dashed) by two standard
deviations. The shaded area is the range of acceptable fits taking
statistical and systematic errors into account. Our results are
consistent with the shift due to the compressional effect given by
Eq. (\ref{shift}), indicated by the solid line. The filled circle
represents the data of Fig.~\ref{findtc}.} \label{tc}
\end{figure}

In Fig.~\ref{tc}, we have plotted $T_{\rm c}$ as a function of
$N_{\rm c}$, measured in ten independent data sets. The ideal gas
value $T_{\rm c}^{\rm ideal}$ (dashed line) lies two standard
deviations above our data. Including the mean-field correction
(\ref{shift}) yields a much better agreement (solid line), that we
quantify in the following way. We assume that the interactive
shift in $T_{\rm c}$ can be written as $\delta T_{\rm c}/T_{\rm
c}^0=\alpha N^{1/6}$, with a free coefficient $\alpha$. A fit to
the data yields $\alpha = -0.009(1)_{-0.001}^{+0.002}$, whereas
Eq.~(\ref{shift}) predicts $\alpha \approx - 0.007$ for a
scattering length $a\approx5.31$ nm \cite{vankempen2002a} and
$\sigma \approx1.00~\mu$m. The first uncertainty quoted is
statistical, while the upper and lower bounds reflect calibration
and analysis uncertainties. The shaded area in Fig.~\ref{tc}
delineates the resulting 1$\sigma$ confidence interval compatible
with the experimental results.

The data shown in Fig.~\ref{tc} reasonably exclude any additional
shift of the same order of magnitude as the compressional effect
given by Eq. (\ref{shift}). In particular, if the (positive)
critical shift in $T_{\rm c}$ predicted in the uniform case
\cite{gruter1997a} were directly scalable to the trapped one, one
would expect an overall $\alpha \approx - 0.004$, a value not
consistent with our findings within the estimated accuracy. This
observation is in line with recent theoretical studies of critical
density fluctuations in a harmonically trapped gas
\cite{arnold2001a,holzmann2003a}, which point out that, instead of
being delocalized over the entire system as in the homogeneous
case, critical fluctuations in the trapped gas are confined to a
small region around the trap center. This reduces corrections to
the equation of state by a factor $\sim(a/\lambda_0)^3$,
corresponding to the ratio of the volume of the fluctuation region
to the volume of the thermal cloud. The critical temperature,
being fixed by the equation of state of the whole cloud, thus
depends only weakly on critical phenomena: corrections to mean
field behavior \cite{secondorder} enter only to second order in
$a/\lambda_0$ \cite{arnold2001a,holzmann2003a}. For our
experimental parameters, we calculate from \cite{arnold2001a} an
upwards correction to $T_{\rm c}$ smaller than 1\%, below the
sensitivity of the measurement.

In summary, we have measured the critical temperature of a
trapped, weakly-interacting $^{87}$Rb Bose-Einstein gas. Our
results exclude ideal gas behavior by two standard deviations, and
we find satisfactory agreement with mean-field theory. We find no
evidence for critical behavior close to $T_{\rm c}$ within our
experimental sensitivity, in line with recent theoretical
estimates that predict an increase of $T_{\rm c}$ due to critical
fluctuations significantly smaller in the trapped case than in the
uniform case \cite{arnold2001a}. We have also observed
hydrodynamic behavior in the expansion of the thermal cloud, and
shown how to correct for it in the thermometry procedure.  We note
to conclude that measuring corrections to $T_{\rm c}$ beyond the
mean-field for our typical experimental parameters would require
thermometry with an accuracy of 1\% or better. A more direct route
to investigate such effects might be to measure the critical
density near the center of the trap, directly sensitive to the
presence of critical fluctuations. Alternatively, these many-body
effects could be enhanced in the vicinity of a Feshbach resonance
\cite{inouye1998a}.

We acknowledge useful discussions with D. Gu{\'e}ry-Odelin
concerning the work reported in \cite{pedri2003a}, and with F.
Lalo{\"e} on critical temperature calculations. We also thank D.
Boiron, J. Retter, J. Dalibard and S. Giorgini for useful comments
on this work. JHT acknowledges support from CNRS, and MH from
IXSEA. This work was supported by DGA, and the European Union.
%


\bibliography{fctc}

\begin{thebibliography}{26}
\bibitem{imaging} The probe laser wavelength is $\lambda_{\rm
L}=780.2$ nm; its intensity is $0.17$\,mW/cm$^2$, and the imaging
pulse length is $34$\,$\mu$s. The half-width of the absorption
resonance has been measured to be 8 MHz, slightly broader than the
natural linewidth 6 MHz.This could explain the small value of the
measured cross-section.
\bibitem{gammacoll}
We have extrapolated the results of \cite{kavoulakis2000a} by the formula $\gamma_{\rm coll}
\approx \gamma_{\rm class}(1+0.23/t^3+0.4/t^4)$, with $t=T/T_{\rm
c}^0$. Following \cite{pedri2003a}, we define the collision rate
for a Boltzmann gas as $\gamma_{\rm class}=(4\sqrt{2\pi}/5) n_{\rm
class}({\bf{0}}) \sigma_{\rm el} v_{0}$, with $\sigma_{\rm
el}=8\pi a^2$ and $n_{\rm class}({\bf{0}})=N
\bar{\omega}^3(m/2\pi k_{\rm B} T)^{3/2}$.
\bibitem{temp}
In practice, we deduce effective temperatures from the measured
sizes $R_{i}$ of the cloud ($i=x,z$), according to $k_{\rm B}
T_{i} = M \omega_{i}^2 R_{i}^2/b_{ i}^2$, with
$b_{i}=(1+\tau_i^2)^{1/2}$ and $\tau_i=\omega_{i}t$. For
$t\gg\omega_\perp^{-1}$ and $t\sim\omega_z^{-1}$, we infer the
initial temperature $T_0$ from $T_{0}\approx 2 \tau_{z}^2/(1+3\tau_{z}^2)
T_{x}+(1+ \tau_{z}^2)/(1+3\tau_{z}^2)T_{z}$, that cancels the
hydrodynamic corrections to first order in $\gamma_{\rm
coll}/\omega_{\perp}$.
\bibitem{secondorder}
Note that these corrections are provided by critical fluctuations
with wavelength comparable or shorter than the correlation length
$\lesssim r_{\rm c}$, which are not qualitatively affected by the
potential.
\end{thebibliography}
\bibliographystyle{apsrev}

\end{document}